\documentclass{aa}  

\usepackage{graphicx}
\usepackage{lscape}
\usepackage{txfonts}
\usepackage{xcolor}
\usepackage{lineno}

\usepackage[normalem]{ulem}

\begin{document} 

\newcommand{\mg}[1]{\textcolor{magenta}{#1}}
\newcommand{\barbara}[1]{{\em B: }{\color{red}{#1}}} 
\newcommand{\cecilia}[1]{{\color{cyan}{C: #1}}} 

    \title{Exploring the connection between fast radio bursts and binary neutron star mergers} 

   \author{B. Patricelli
          \inst{1,2,3}
          \and
           M.G. Bernardini
           \inst{4}
           \and C. Sgalletta
           \inst{5}
           \and
           M. Mapelli
           \inst{5,6,7,8}
          }

   \institute{Physics Department, University of Pisa,
              Largo B. Pontecorvo 3, I-56127 Pisa, Italy\\
              \email{barbara.patricelli@pi.infn.it}
         \and
             INFN - Pisa, Largo B. Pontecorvo 3, I-56127 Pisa, Italy
         \and
            INAF - Osservatorio Astronomico di Roma, Via Frascati 33, I-00078 Monte Porzio Catone (Rome), Italy
         \and
            INAF - Osservatorio Astronomico di Brera, via Bianchi 46, I-23807 Merate (LC), Italy
         \and 
              Universit\"at Heidelberg, Zentrum f\"ur Astronomie (ZAH), Institut f\"ur Theoretische Astrophysik, Albert-Ueberle-Str. 2, 69120,\\ $^{\,}$ Heidelberg, Germany
              \and
            Universit\"at Heidelberg, Interdisziplin\"ares Zentrum f\"ur Wissenschaftliches Rechnen, Heidelberg, Germany
            \and
            Physics and Astronomy Department Galileo Galilei, University of Padova, Vicolo dell'Osservatorio 3, I--35122, Padova, Italy
            \and
    INFN, Sezione di Padova, Via Marzolo 8, I--35131 Padova, Italy
         }

\date{}

  \abstract 
   {Fast radio bursts (FRBs) are highly energetic radio sources with millisecond-scale durations. The physical origin of these sources remains unknown. Many models suggest magnetars as possible progenitors of FRBs, supported by the association between FRBs and the Galactic magnetar SGR 1935+2154. Other proposed progenitors include binary neutron star (BNS) mergers, which are themselves linked to magnetar formation.}
   {In this work we investigate the possible connection between FRBs and BNS mergers -- including  magnetars  which may be produced in such events -- by comparing the detection rates inferred from synthetic BNS and associated FRB populations with those observed by CHIME.}
   {We produced a synthetic catalogue of BNS mergers by combining recent theoretically predicted BNS merger rates as functions of redshift and the neutron star mass distribution inferred from measurements of Galactic BNSs. Using this catalogue we predicted the number of BNS systems that result in magnetars (stable or supramassive neutron stars) or black holes (formed promptly or via hypermassive neutron star collapse) for different equations of state. For each BNS (and therefore for each magnetar remnant), we then simulated an associated FRB and estimated the fraction potentially detectable by CHIME.}
   {We find that the rates of BNS mergers and magnetars produced by BNS mergers represent a non-negligible fraction of the FRBs detected by CHIME, including both repeating and non-repeating events.}
   {Although additional formation channels must be considered to account for the entire population of FRBs, the existence of a fraction of FRBs that may genuinely be associated with BNS mergers has profound implications for multi-messenger astronomy, supporting the systematic searches for coincident detections of FRBs and gravitation waves from a BNS merger with current and future facilities.}


   \maketitle

\section{Introduction}

Fast radio bursts (FRBs; see \citealp{2020Natur.587...45Z} for a review)  are millisecond-duration flashes of radio waves from extragalactic sources \citep{2007Sci...318..777L,2013Sci...341...53T}. To
date, thousands of FRBs have been detected at frequencies ranging from 400 MHz to 8 GHz by several ground-based radio facilities (see e.g. \citealp{2026ApJS..283...34C}). They are observationally classified into two categories: non-repeating or repeating sources (see e.g. \citealp{2016Natur.531..202S,2026ApJS..283...34C}).  

The physical origin of FRBs remains unknown, and many FRB progenitor models have been proposed in recent years (see e.g. \citealp{2019PhR...821....1P}). Such models can be grouped into two main categories: non-catastrophic models (mainly related to repeating FRBs) and catastrophic models (mainly related to non-repeating FRBs). Coalescing compact objects represent one of the proposed progenitor channels for non-repeating FRBs. Specifically, several  studies have investigated the possibility of FRB-like emission from
binary neutron star (BNS) mergers (see e.g. \citealp{2020ApJ...890L..24Z}). Other models consider magnetars as possible sources of FRBs. 

The magnetar origin of at least some FRBs has been confirmed by the recent observation of a bright radio burst reported by the \cite{2020Natur.587...54C} and by the STARE2 radio array \citep{2020Natur.587...59B}. This burst is spatially and temporally coincident with an X-ray burst from the Galactic magnetar SGR 1935+2154, observed with INTEGRAL \citep{2020ApJ...898L..29M}, AGILE \citep{2021NatAs...5..401T}, Konus-Wind \citep{2021NatAs...5..372R}, and Insight-HXMT \citep{2021NatAs...5..378L}. Additionally, further observational evidence supports this scenario. For instance, the properties of the
 host galaxy of the first repeating FRB, FRB 121102 \citep{2017ApJ...834L...7T,2017ApJ...843L...8B}, and the discovery of a persistent radio synchrotron source spatially coincident with it \citep{2017ApJ...834L...8M} suggest a young magnetar born from a catastrophic event, such as a superluminous supernova or a long-duration gamma-ray burst (GRB) \citep{2017ApJ...841...14M,2017ApJ...843...84N}.

One of the possible formation channels of magnetars is the coalescence of BNS systems. In fact, depending on the masses of the two neutron stars (NSs) and on the equation of state (EOS) of nuclear matter (see e.g. \citealp{2006PhRvD..73f4027S,2008PhRvD..78h4033B,2011PhRvD..83l4008H}), the  BNS merger can form: i) a black hole (BH) from prompt collapse; ii) a hypermassive neutron star (HMNS), which collapses into a BH on a timescale of the order of milliseconds to $\sim$ 100 ms; iii) a supramassive neutron star (SMNS), which collapses into a BH on a timescale of the order of seconds, minutes, or longer; and iv) a stable NS.  

Here, we explore the possible connection between FRBs and BNSs, investigating both BNS mergers and magnetars produced in BNS mergers as possible progenitors. We estimated the rate of detectable FRBs assuming that they are associated either with BNS mergers or their magnetar remnants, and we compared this rate with that of the FRBs observed by the Canadian Hydrogen Intensity Mapping Experiment (CHIME). The work is organized as follows. In Sect. \ref{sec:BNSpop} we explain how the synthetic catalogues of BNS mergers were produced, while in Sect. \ref{sec:FRBpop} we describe in detail how the associated populations of FRBs were simulated. In Sect. \ref{sec:CHIME-FRBs} we present our FRB detection rates and compare them with that of the FRBs observed by CHIME. Finally, in Sect. \ref{sec:results} we discuss our results and summarize our conclusions.

\section{The BNS population}\label{sec:BNSpop}

We generated a sample of synthetic BNS mergers according to a theoretically predicted cosmic BNS merger rate (see Sect. \ref{sec:cosmicBNSmr}). We assumed an isotropic and
homogeneous distribution in space for the simulated BNSs. 
The remnant of the simulated 
BNS mergers (a NS or a BH) was lastly estimated considering different EOSs and accounting for the masses of the two NSs in the binary (see Sect. \ref{sec:remnant}).

\subsection{ The cosmic BNS merger rate}\label{sec:cosmicBNSmr}

We generated catalogues of merging BNSs with the `Stellar Evolution for N-body' (\textsc{sevn}; \citealt{Spera2017, Spera2019, Mapelli2020, Iorio2023}) code. \textsc{sevn} is a binary population synthesis code that evolves stars by interpolating on-the-fly, pre-computed stellar tracks. Additionally, \textsc{sevn} implements binary interactions through analytical and semi-analytical formulas.
For the purpose of this work, we adopted the fiducial \textsc{sevn} setup, as described by \citet{Iorio2023}. Here, we used the BNS evolutionary model described by \cite{Sgalletta2023}, which includes pulsar spin down, recycling, and the propeller effect. In particular, we modelled the output of core-collapse supernovae by assuming the delayed prescription by \citet{Fryer2012}. We estimated natal kicks following \citet{2020ApJ...891..141G}. According to this model, the natal kick velocity $v_{\rm kick}$ is derived as
\begin{equation}
    v_{\rm kick} = f_{\rm H05} \frac{M_{\rm ej}}{\langle M_{\rm ej}\rangle} \frac{\langle M_{\rm NS}\rangle}{M_{\rm rem}},
\end{equation}
where $f_{\rm H05}$ is a random number drawn from a Maxwellian distribution with a root mean square $\sigma = 217$ km s$^{-1}$ \citep{Disberg2025}, $M_{\rm ej}$ is the ejected mass during the explosion, and $ M_{\rm rem}$ is the mass of the remnant. 
The constants $\langle M_{\rm ej}\rangle$ and $\langle M_{\rm NS}\rangle$ are the averages, evaluated over a population of single stars, of the ejected mass associated with a NS formation and of the NS mass, respectively.

We modelled mass transfer, common-envelope evolution, tidal evolution, and gravitational-wave decay, as described by \citet{Hurley2002} and \citet{Iorio2023}. Adopting the model QCRS by \citet{Iorio2023} and \citet{Sgalletta2023}, we assumed, in particular, that the mass transfer with donor stars in the main sequence and the Hertzsprung gap remains stable. We ran three sets of simulations varying $\alpha=0.1$, $0.5$, and $0.7$. During a common envelope event, $\alpha$ parametrizes the efficiency at which orbital energy is transferred to the envelope. Such energy is used by the binary to unbind the envelope. We considered this range in $\alpha$ to bracket the uncertainties on the BNS merger rate, yielding local BNS merger rates of $3.4$, $53.0$, and $156.0$ Gpc$^{-3}$ yr$^{-1}$ for $\alpha=0.1$, 0.5, and 0.7, respectively. These rates encompass the BNS merger rate inferred from the latest LIGO--Virgo--KAGRA observations \citep[GWTC-5;][]{2026arXiv260527226T}.
We did not consider models with higher $\alpha$ values as they result in higher BNS merger rates \citep[see e.g.][]{Sgalletta2023, Iorio2023}, which exceed the 90\% credible interval inferred from GWTC-5 data \citep{2026arXiv260527226T}.

For every value of $\alpha$, we generated and evolved $2 \times 10^6$ binaries for 12 metallicities: $Z=0.0002$, $0.0004$, $0.0008$, $0.0012$, $0.0016$, $0.002$, $0.004$, $0.006$, $0.008$, $0.012$, $0.016$, $0.02$. We sampled the primary masses of the progenitor binary stars in the range $M_1 \in [5,150]$ M$_\odot$ from a Kroupa initial mass function \citep{Kroupa2001}. We sampled the mass ratio according to the fit by \citet{Sana2012}: $q=M_2/M_1$ from $\mathcal{F}(q) \propto q^{-0.1}$ with $q \in [\min(0.1, 2.2 \text{M}_\odot/ M_1), 1]$. We abided by the fit in \citet{Sana2012} to sample the distributions of periods and eccentricities with $\mathcal{F}(P_{\rm orb}) \propto (\log P_{\rm orb})^{-0.55}$, $\log (P_{\rm orb}/\text{day})\in [0.15, 5.5]$ and $\mathcal{F}(e) \propto e^{-0.42}$, $e\in [0, 1-\left(P/2 \ \mathrm{days}\right)^{-2/3}]$. Here, we included the correction by \citet{Moe2017}.

We evaluated the merger rate density of BNSs, $\mathcal{R}_{\rm BNS}(z)$, using galaxy$\mathcal{R}$ate \citep{Santoliquido2022,Sgalletta2025}:
\begin{equation} \label{eq:mrd}
    \mathcal{R}_{\rm BNS}(z) = \frac{1}{V^3} \int_{z_{\rm max}}^{z} 
    \left[ \int_{Z_{\rm min}}^{Z_{\rm max}} \mathcal{S} \left( z', Z \right) \mathcal{F}_{\rm BNS}\left( z', z, Z \right) dZ \right] \frac{dt(z')}{dz'} dz',
\end{equation}
where $V \sim ( 100$ cMpc$)^3$ is the total simulated co-moving volume, $z_{\rm max}=8$ is the maximum formation redshift of the stellar progenitors, and $\mathcal{S}(z', Z)$ is the star formation rate density. Here, $\mathcal{F}_{\rm BNS}\left( z', z, Z \right)$ is defined as
\begin{equation} \label{eq:catalogs}
    \mathcal{F}\left( z', z, Z \right) = \frac{1}{M_{\rm sim}} \frac{\mathcal{N} \left( z', z, Z \right)}{dt} f_{\rm bin} f_{\rm corr},
\end{equation}
where $M_{\rm sim}$ is the total simulated initial mass, $\mathcal{N}_{\rm BNS} \left( z', z, Z \right) \ dt$ is the rate of BNSs forming at $z'$, with metallicity $Z$, and merge at redshift $z$. 
The factors $f_{\rm bin}=0.5$ and $f_{\rm corr} =0.251$ account for the fraction of binaries and the incomplete sampling of the initial mass function, respectively. 
galaxy$\mathcal{R}$ate exploits data-driven observational scaling relations to generate a set of star-forming galaxies distributed across redshifts. The star formation rate density $\mathcal{S} \left( z', Z \right)$ results from integrating the star formation rates of the galaxies in the set. In this work, we adopted the same fiducial galaxy$\mathcal{R}$ate settings as in \citet{Sgalletta2025}, which include the most up-to-date observational scaling relations.

The rate of events within $z\leq 8$ was estimated as follows. The co-moving volume element is 
\begin{equation}
dV(z)=\frac{c}{H_0} \frac{D_c^2}{E(z)} d\Omega dz,  
\end{equation}
where c is the speed of light, $d\Omega$ is the solid angle,  \mbox{$E(z)=\sqrt{\Omega_{\rm{M}} (1+z)^3+\Omega_{\rm{K}}  (1+z)^2+\Omega_\Lambda}$} , and $D_c$ is the co-moving distance expressed as  
\begin{equation}
D_c(z)=\frac{c}{H_0} \int_0^z{\frac{dz'}{E(z')}}.    
\end{equation}
We used the cosmological parameters from \cite{2020A&A...641A...6P}. 

The total BNS merger rate within a redshift z in the observer frame is given by
\begin{equation}
n_{\rm obs}(< z)=4 \pi \int_0^z{\frac{R_{\rm BNS}(z')}{(1+z')}\frac{dV(z')}{dz'}dz'} \quad [yr^{-1}].
\end{equation}

\subsection{The NS mass distribution}\label{sec:massdistr}

Neutron star masses have been estimated in the past through both electromagnetic (EM) and gravitational wave (GW) observations. The least and most massive NSs observed to date through EM observations are the companion of the pulsar J0453+1559, with a mass M=1.174 $\pm$ 0.004 M$_\odot$ \citep{2015ApJ...812..143M}, and the millisecond pulsar (MSP) J0740+6620, with an estimated mass M=$2.14^{+0.10}_{-0.09}$ M$_\odot$ (68\% credible interval; \citealp{2020NatAs...4...72C}).
Through GWs, only two BNS systems have been observed so far: GW170817 \citep{2017PhRvL.119p1101A} and GW190425 \citep{2020ApJ...892L...3A}. The masses of the NSs in these two binary systems lie within the range of masses for Galactic BNSs under the `low-spin' assumption (i.e. spins restricted to
the range observed in Galactic BNSs). For GW190425, a higher mass for one of the two components is allowed when considering higher spins, with the mass of the primary component in the range 
1.61--2.52 M$_\odot$ (see \citealp{2020ApJ...892L...3A}). 

Following the approach used by \cite{2020MNRAS.499L..96P},  we assumed that both the NSs in the binary 
systems have a mass distribution equal to that inferred from the measurements related to the Galactic binary systems: a Gaussian distribution with a central mass of 1.33 M$_\odot$ and a dispersion of 0.09 M$_\odot$
\citep{2016ARA&A..54..401O}. We assumed the two Gaussians to be uncorrelated. We also assumed that the mass distribution does not depend on redshift (see e.g. \citealp{2019MNRAS.487....2M,2019MNRAS.482..870E}).

\subsection{The EOS and the BNS merger outcome}\label{sec:remnant}

In addition to the masses of the two NSs in the binary system, the other key element required to identify the nature of the remnant of a BNS merger is the EOS. Although it has been widely studied for decades, the EOS of cold, ultra-dense matter at high densities remains poorly constrained (see e.g. \citealp{2001ApJ...550..426L}). Some constraints on the EOS can be put through GW observations. In fact, GWs from binary inspirals are influenced by the tidal deformation that a star’s gravitational field induces on its companion.
One example of this is GW170817 \citep{2018PhRvL.121p1101A}. Its observation shows that `soft' EOSs, such as APR4 \citep{1998PhRvC..58.1804A}, which predict lower values of the tidal deformability parameter, are favoured over `stiff' EOSs, such as H4 \citep{1991PhRvL..67.2414G} and MS1 \citep{1996NuPhA.606..508M}. These latter EOSs predict higher values of the tidal deformability parameter and lie outside the 90\% credible region. Other constraints have been derived from NICER's mass-radius estimate for PSR J0740+6620 and multi-messenger observations of GW170817 \citep{2021ApJ...918L..29R}. 

In this work, we adopted the following EOSs: APR4, MS1, and H4. These three EOSs cover a relatively wide range of maximum NS masses but with a maximum gravitational mass $\gtrsim$ 2 M$_\odot$, consistent with the current observational EM limits. 

To estimate the remnant produced by the BNS mergers, we followed the approach proposed by \cite{2017ApJ...844L..19P} and used in \cite{2020MNRAS.499L..96P}. First, we considered a reference value equal to M$_{\rm lost}$=0.01 M$_\odot$ for the mass lost from the system during the merger. Then, we converted the gravitational masses of the simulated NSs to baryonic masses (m$_{\rm{b,1}}$ and m$_{\rm{b,2}}$) using the model developed by \cite{2017ApJ...844L..19P} for non-rotating NS\footnote{\cite{2017ApJ...844L..19P} note that, during the inspiral phase, the two NSs are not strongly affected by tidal coupling. Thus, they are not significantly spun up, and their structure is well approximated by non-spinning NS models.} and for the three different EOSs. Finally, we estimated the total baryonic mass of the remnant as
\begin{equation}
{\rm M_{b,tot}}={\rm m_{b,1}}+{\rm m_{b,2}}-{\rm M_{\rm lost}}.
\end{equation}
The value of M$_{\rm{b,tot}}$ was then compared with the maximum NS baryonic masses to evaluate whether the remnant is a NS or a BH (see Table 1 of \citealp{2017ApJ...844L..19P}). Remnants that are stable NSs or SMNSs were considered as `magnetar-like' outcomes.

\section{The FRB population}\label{sec:FRBpop}

We began from the assumption that every BNS merger/magnetar remnant is associated with an FRB and estimated the detection rate of this population of FRBs. To do this, we generated a synthetic population of FRBs,
assigning to each: i) the same redshift and sky coordinates of the simulated BNSs and ii) a rest-frame isotropic energy, drawn from the FRB
energy distribution $\Phi(E)$. 
The FRB luminosity and energy distributions are typically described with a Schechter function 
\citep{1976ApJ...203..297S}, whose parameters are constrained by observations (see e.g. \citealp{2019ApJ...883...40L,2020MNRAS.494..665L,2022MNRAS.511.1961H}). Following the approach used by \cite{2024A&A...689A.286P}, we used the energy distribution derived by \cite{2022MNRAS.511.1961H} (hereafter: H22) and considered repeating and non-repeating FRBs separately.  

H22 used a homogeneous sub-sample of FRBs from the first CHIME catalogue. The FRBs were first classified as repeating and non-repeating and then, depending on their redshift, further divided into several sub-samples, filling different redshift bins, covering the range from 0.05 to 3.6 for non-repeating FRBs and from 0.05 to 1.5 for repeating FRBs. Since results may depend on how the redshift bins are selected, H22 performed their analysis considering two different sets of redshift bins (labelled `redshift A' and `redshift B'). They fit Schechter functions to the derived energy functions and accounted for the redshift evolution. They report $\alpha$=-1.4$^{+0.7}_{-0.5}$ (-1.1$^{+0.6}_{-0.4}$) for redshift bin A (redshift bin B).

Since the redshift range of BNSs is wider than that considered in H22, we assumed that the energy distribution $\phi(E)$ of FRBs with z < 0.05 (> 3.6/1.5) is the same as that obtained for the lowest (highest) redshift bin. We considered both the redshift A and the redshift B case, adopting the parameters reported in Table 1 of H22. We then associated each FRB with a rest-frame isotropic energy $E_{\rm{rest,400}}$ integrated over the CHIME frequency width $\Delta\nu=$400 MHz. We assumed a fiducial value of 10$^{37}$ ergs (i.e. the minimum energy of the FRBs in the CHIME catalogue) for the minimum FRB rest-frame energy and a maximum value of 10$^{50}$ ergs. 

From these values, we computed the observed fluence, $F_{\nu}$, in the CHIME frequency band (400 MHz - 800 MHz) as 
\begin{equation}
{\rm F_\nu=\frac{(1+z)^{2-\gamma}\,{} E_{\rm{rest,400}}}{4 \pi d_L^2(z) \Delta\nu},}
\label{eq:Fnu}
\end{equation}
where $d_L(z)$ is the luminosity distance at redshift z, estimated using the cosmological parameters from \citet{2020A&A...641A...6P}; $\Delta\nu$ is the frequency bandwidth (taken as 400 MHz); and $\gamma$ is the FRB spectral index (F $\propto \nu^{-\gamma}$), assumed to be equal to 1.4 (see e.g. \citealp{2023ApJ...944..105S}). 

\section{The detection rate of CHIME FRBs}\label{sec:CHIME-FRBs}

We consider an FRB from our parent population as detectable if its $F_\nu$ (from Eq. \ref{eq:Fnu}) lies above the CHIME completeness threshold (95$\%$ c.l.) of ${\rm F_{lim}=3.5}$ Jy ms to account for the different sources of sensitivity variation \citep{2026ApJS..283...34C}. We thus computed the rate of detected FRBs by accounting for the CHIME instantaneous field of view (fov) of $120^\circ\times 2^\circ$ and its duty cycle of 76 \% \citep{2026ApJS..283...34C}.

The detection rates for non-repeating FRBs from different parent populations are reported in Table \ref{tab:SimDetRates-nrFrb}. This table lists FRBs associated with magnetars (stable NSs plus SMNSs) as well as FRBs originating from the BNS merger events, regardless of the remnant. 
It can be seen that the rates of detectable FRBs are very sensitive to both the common envelope efficiency, which directly influences the rate of BNS mergers (as discussed in Sect. \ref{sec:BNSpop}), and the EOS, which determines the fraction of BNS mergers resulting in magnetars.  
The rate of detectable non-repeating FRBs originating from magnetars is lowest for the H4 EOS and highest for the MS1 EOS, accounting for approximately 14\% and 100\%, respectively, of the total rate of detectable FRBs originating from BNS mergers (regardless of the remnant).

For comparison, the most updated catalogue of FRBs detected by the CHIME telescope reports an observed number of $1.91\pm 0.04$ sources per day \citep{2026ApJS..283...34C}, namely a rate of $698$ yr$^{-1}$. By comparing the simulated FRB detection rates with the observed CHIME FRB detection rates for non-repeating sources, it can be seen that BNS mergers and magnetars resulting from BNS mergers could contribute up to $\sim$ 2 \% of the total rate of observed FRBs. It is worth noting that, as mentioned above, the rate of total BNS mergers producing a detectable FRB is equal to the rate of magnetars producing a detectable FRB when the MS1 EOS is considered. In fact, for such an EOS the rate of merging BNS systems resulting in a BH or an HMNS is negligible.
 
\begin{table}[h!]
\caption{Non-repeating FRB detection rates.}
\begin{center}
\begin{tabular}{|c|ccc|c|} 
\hline\hline
$\alpha$ & \multicolumn{3}{|c|}{Stable NS + SMNS} & Total BNS\\
\hline
         & MS1& APR4 & H4 & \\
& yr$^{-1}$ & yr$^{-1}$ & yr$^{-1}$ & yr$^{-1}$\\
\hline         
\hline
0.1 & 0.01 - 0.03 & 0.01 - 0.03 & 0.00 - 0.00 & 0.01 - 0.03\\
0.5 & 1.11 - 4.34 & 0.90 - 3.57 & 0.15 - 0.62 & 1.11- 4.34 \\
0.7 & 3.65 - 14.92 & 2.98 - 12.08 & 0.58 - 2.07 & 3.65 - 14.92 \\

 \hline
\end{tabular}
\tablefoot{Non-repeating FRB detection rates, considering the different BNS models, EOSs and redshift bins. For the various EOSs, the rate of FRBs associated with both stable NSs and SMNSs is reported. The lower and upper boundaries correspond to redshift bin A and B, respectively.}
\end{center}
\label{tab:SimDetRates-nrFrb} 
\end{table}

We compared the BNS mergers with the repeating FRBs observed by CHIME. Since repeating FRBs are not  associated with cataclysmic events, we considered only BNSs resulting in stable magnetars. As for the non-repeating FRBs, we assumed that each magnetar produces at least one FRB and selected those whose fluxes lie  above the CHIME completeness threshold during their lifetime. We then computed the number of newly discovered FRB sources from this magnetar population. For different EOSs and values of $\alpha$, these numbers are reported in Table \ref{tab:SimDetRates-rFRB} and should be considered as upper limits when qualitatively compared to the actual individual repeating FRB sources discovered by CHIME in its second catalogue (30 in 1878 days, between 2018 July 25 and 2023 September 15; \citealt{2026arXiv260508410C}). In fact, we required that at least one burst from each source lie above the detection threshold, whereas the secure identification of a repeating FRB source requires the detection of multiple bursts.

\begin{table}[h!]
\caption{Repeating FRB detection rates.}
\begin{center}
\begin{tabular}{|c|ccc|} 
\hline\hline
$\alpha$ & \multicolumn{3}{|c|}{Stable NS} \\
\hline
         & MS1& APR4 & H4 \\
& yr$^{-1}$ & yr$^{-1}$ & yr$^{-1}$ \\
\hline
0.1 & 0.000 - 0.013 & 0.000 - 0.000 & 0.000 - 0.000\\
0.5 & 0.044 - 1.353 & 0.004 - 0.027 & 0.000 - 0.000 \\
0.7 & 0.195 - 4.116 & 0.000 -  0.097 & 0.000 - 0.000\\
 \hline
\end{tabular}
\tablefoot{Rate of newly discovered candidate repeating FRB sources detected by the CHIME survey, considering the different BNS models, EOSs, and redshift bins.}
\end{center}
\label{tab:SimDetRates-rFRB} 
\end{table}

\section{Discussion and conclusions}\label{sec:results}

In Sect. \ref{sec:CHIME-FRBs} we showed that a non-negligible fraction of non-repeating FRBs observed by CHIME may indeed be produced by BNS merger events. Our results are limited to the remnants of merging BNS systems formed through isolated binary evolution, which are believed to dominate the total BNS merger rates (see e.g. \citealp{2020ApJ...888L..10Y,2020ApJ...898..152S}). The contribution of systems formed through dynamical assembly is not expected to change the conclusions of this paper. 

Constraining the parent population of repeating FRBs requires knowledge of their birth rates, which are very loose ($1-10^6$ Gpc$^{-3}$ yr$^{-1}$, \citealp{2023PASA...40...57J}; $2.2\times 10^2-5.2\times 10^4$, \citealp{2022ApJ...927...55L}). These ranges are compatible with most FRB progenitor models. In this work, we provided an order-of-magnitude estimate of candidate repeating FRB sources that could have been discovered by the CHIME survey as a result of BNS mergers. As for the case of non-repeating FRBs, stable magnetars from BNS mergers can account for a non-negligible fraction of the repeating FRB sources form the CHIME second catalogue.

Indeed, there is evidence supporting the association of several FRBs with catastrophic events such as superluminous supernovae (SNe) and long GRBs  \citep{2017ApJ...841...14M,2017ApJ...843...84N}. Examples include FRB 121102 \citep{2017ApJ...834L...7T,2017ApJ...843L...8B}, FRB 20180916B \citep{2020Natur.577..190M,2021ApJ...908L..12T}, and FRB 20201124A \citep{2022ApJ...927L...3N}. Thus, magnetars produced from progenitors with higher local rates than BNS mergers, such as core-collapse SNe ($10^5$ Gpc$^{-3}$ yr$^{-1}$, \citealp{2015A&A...584A..62C,2020ApJ...904...35P}), may dominate the observed population of FRBs. However, even in this case, a population of more exotic magnetars than those produced in core-collapse SNe seems required to account for the entire population of extragalactic FRBs \citep{2020ApJ...899L..27M,2025A&A...700A..19G}.

The existence of a fraction of FRBs that may be genuinely associated with BNS mergers has profound implications for multi-messenger astronomy. Indeed, a definitive probe of this association could originate from a coincident detection of FRBs and GWs from BNS mergers. Several searches for GW counterparts to FRBs (including BNS mergers) have been performed in the past. Although no significant association has been found so far (see e.g. \citealp{2023ApJ...955..155A,2024ApJ...977..255A}), our results highlight the importance of performing systematic searches to validate this potential connection. In the future, the increasing sensitivity of GW detectors\footnote{https://observing.docs.ligo.org/plan/} and the start of operations of new radio detectors may enhance the prospects for such multi-messenger detections. For example, the advent of the Canadian Hydrogen Observatory and Radio-transient Detector (CHORD; \citealp{2019clrp.2020...28V}) -- a new radio telescope facility an order of magnitude more powerful than CHIME -- is expected to provide a significant boost in the determination of FRB progenitors. CHORD will provide routinely sub-arc-second localizations, enabling systematic identification of FRB host galaxies. Other upcoming radio telescopes expected to significantly improve our understanding of FRBs include: the Bustling Universe Radio Survey Telescope in Taiwan (BURSTT;  \citealp{2022PASP..134i4106L}), optimized to discover and localize a large sample of rare, high-fluence, and nearby FRBs; the Deep Synoptic Array 2000-antenna concept (DSA-2000;  \citealp{2019BAAS...51g.255H}) which, similar to CHORD, will increase the FRB discovery rate to at least dozens of sources per day while providing sub-arc-second sky localizations; and the Square Kilometre Array (SKA; \citealp{2015aska.confE..51F}), which will operate with greater sensitivity (by an order of magnitude) and survey speed than any current radio telescope.

\begin{acknowledgements}
We thank Maura Pilia for useful discussions. We acknowledge support from the Deutsche Forschungsgemeinschaft (DFG, German Research Foundation) through project number 546850815 (acronym: DoBlack) and under Germany's Excellence Strategy EXC 2181/1 - 390900948 (the Heidelberg STRUCTURES Excellence Cluster). CS acknowledges financial support from the Alexander von Humboldt Foundation for the Humboldt Research Fellowship. MGB and BP acknowledge support from the INAF grant no. 1.05.24.02.23.
The BNS simulations were performed thanks to the support of the state of Baden-W\"urttemberg through bwHPC and the DFG through grants INST 35/1597-1 FUGG and INST 35/1503-1 FUGG. 
\end{acknowledgements}

\section*{Data availability}
The data underlying this article will be shared on reasonable request to the corresponding author.

\bibliographystyle{aa}
\bibliography{bibliography}

\end{document}